\documentclass[sigconf]{acmart}
\usepackage[ruled]{algorithm2e}
\SetKwComment{Comment}{/* }{ */}
\usepackage{multirow}

\AtBeginDocument{%
  }

\copyrightyear{2024}
\acmYear{2024}
\setcopyright{acmlicensed}
\acmConference[HPDC '24]{The 33rd International Symposium on High-Performance Parallel and Distributed Computing}{June 3--7, 2024}{Pisa, Italy}
\acmBooktitle{The 33rd International Symposium on High-Performance Parallel and Distributed Computing (HPDC '24), June 3--7, 2024, Pisa, Italy}
\acmDOI{10.1145/3625549.3658657}
\acmISBN{979-8-4007-0413-0/24/06}

\begin{document}

\title{ESG: Pipeline-Conscious Efficient Scheduling of DNN Workflows on Serverless Platforms with Shareable GPUs}

\author{Xinning Hui}
\email{xhui@ncsu.edu}
\affiliation{%
  \institution{North Carolina State University}
  \city{Raleigh}
  \state{NC}
  \country{USA}
  \postcode{27606}
}

\author{Yuanchao Xu}
\email{yxu314@ucsc.edu}
\affiliation{%
  \institution{University of California, Santa Cruz}
  \city{Santa Cruz}
  \country{USA}}

\author{Zhishan Guo}
\email{zguo32@ncsu.edu}
\affiliation{%
  \institution{North Carolina State University}
  \city{Raleigh}
  \state{NC}
  \country{USA}
}

\author{Xipeng Shen}
\email{xshen5@ncsu.edu}
\affiliation{%
 \institution{North Carolina State University}
 \streetaddress{Rono-Hills}
 \city{Raleigh}
 \state{NC}
 \country{USA}}

\renewcommand{\shortauthors}{Hui et al.}

\begin{abstract}
Recent years have witnessed increasing interest in machine learning inferences on serverless computing for its auto-scaling and cost effective properties. Existing serverless computing, however, lacks effective job scheduling methods to handle the schedule space dramatically expanded by GPU sharing, task batching, and inter-task relations. Prior solutions have dodged the issue by neglecting some important factors, leaving some large performance potential locked. This paper presents ESG, a new scheduling algorithm that directly addresses the difficulties. 
ESG treats sharable GPU as a first-order factor in scheduling. It employs an {\em optimality-guided adaptive} method by combining A*-search and a novel {\em dual-blade pruning} to dramatically prune the scheduling space without compromising the quality. It further introduces a novel method, {\em dominator-based SLO distribution}, to ensure the scalability of the scheduler. The results show that ESG can significantly improve the SLO hit rates (61\%-80\%) while saving 47\%-187\% costs over prior work.
\end{abstract}




\received{25 January 2024}
\received[accepted]{25 March 2024}

\maketitle

\section{Introduction}
\label{sec:intro}






Recent years have witnessed a rapidly growing interest in machine learning (ML) inferences on serverless platforms, thanks to the ease of programming and maintenance, autoscaling, and pay-as-you-go billing of serverless computing~\cite{eismann2020review,wang2019distributed,carreira2019cirrus,yang2022infless, ao2018sprocket,yi2017lavea}. 


Due to the computing-intensive nature of ML inferences, it is desirable to provide GPU support on serverless platforms to ML workloads, which can help significantly boost the service throughput for the massive parallelism of GPUs~\cite{dhakal2020gslice, choi2022serving}. However, the current state of commercial serverless platforms (e.g., AWS Lambda~\cite{awslambda}, Google Cloud Function~\cite{googlelambda}, Azure Function~\cite{azurefunction}, and Knative~\cite{knative}) is falling behind: The schedulers and resource management are still CPU-centric, oblivious to GPU resources. 

There have been some recent research efforts toward adding GPU support to ML on serverless platforms. Some approaches have leveraged NVIDIA MPS~\cite{mps} to facilitate GPU sharing across distinct function instances~\cite{yang2022infless, gu2023fast, cho2022sla}. Others have introduced techniques to enable batching for ML inference, thus increasing overall throughput~\cite{ali2020batch, yang2022infless}. Additionally, certain studies have extended scheduling algorithms to manage heterogeneous computing resources more effectively~\cite{du2022serverless, vandebon2021scheduling}.

Despite the progress made so far, two challenges remain, which have kept much of the potential of GPU out of reach for serverless ML. 

(i) {\bf The dramatically expanded search space for scheduling.}
An important step in serverless scheduling is to {\em configure the serverless functions}, that is to determine the amount of resource to assign to each of the serverless functions in an application to meet the Service Level Objective (SLO) while consuming the minimum resource. The configuration space is as large as $n=m^k$, where $m$ is the number of resource allocation options for one serverless function in the application, and $k$ is the number of functions in the application. Without considering sharable GPUs, the options for a function are just the number of vCPUs to consider. With sharable GPUs, the space becomes three-dimensional: {\em (batch size, number of vCPUs, number of vGPUs)}. The "batch size" here refers to the number of invocations of a serverless function to be grouped into one batch to run. Batching is essential for GPU throughput due to its massive parallelism. The search space hence expands cubically, from $n=m^k$ to $n=(m^k)^3$ (assuming the same number of options in all dimensions and a function can have its own options values). For $m=5, k=7$, the space increases from 78K options to 476 trillions. 

(ii) {\bf The significantly increased complexity in handling performance variations.} Because the number of different states of available resources is multiplied by the available vGPUs, it calls for an adaptive scheduling algorithm to handle them properly. Performance variation is an inherent problem in serverless computing, where the running times of a serverless function vary much across invocations~\cite{mahgoub2022orion}. Batching exacerbates it as unpredictable job arrival times cause variations in the waiting time for forming a batch. Along with these variations, the system workload and resource availability on a serverless system show constant changes. Therefore, adapting agilely to dynamic changes and variations becomes crucial for a scheduler. 

\begin{table}[]
\centering
\scriptsize
\caption{Comparison of serverless systems}
\begin{tabular}{|c|c|c|c|c|c|}
\hline
\textbf{Features }   &\textbf{INFless}   & \textbf{Fast-GShare}  & \textbf{Orion}      & \textbf{Aquatope} &\textbf{ESG}  \\ \hline
\textbf{GPU sharing}             &   $\surd$         &   $\surd$    &  $\times$     &   $\times$   &       $\surd$     \\ \hline
\textbf{\begin{tabular}[c]{@{}c@{}}Inter-function \\relation\end{tabular}} &   $\times$        &   $\times$   &  $\surd$      &  $\surd$    &       $\surd$       \\ \hline
\textbf{Adaptive sched.}         &   $\surd$         &   $\surd$    &  $\times$     &  $\times$    &      $\surd$     \\ \hline
\textbf{Data locality}           &   $\times$        &   $\times$   &  $\times$     &   $\times$   &        $\surd$    \\ \hline
\textbf{Pre-warming}             &   $\surd$         &   $\times$   &  $\surd$      &   $\surd$   &        $\surd$     \\ \hline
\end{tabular}
\label{tab:comparison}
\end{table}

Previous studies have all dodged those challenges by neglecting some important dimensions of the problem (Table~\ref{tab:comparison}), causing a substantial loss in the quality of scheduling. These dimensions include {\bf inter-function relations}, {\bf GPU sharing}, and {\bf runtime system state variations}. Specifically, based on the neglected dimension(s), previous solutions fall into one of two groups. (i) Those works {\em neglecting the relations between functions}, represented by Fast-GShare~\cite{gu2023fast} and INFless~\cite{yang2022infless}, two most recent studies on sharable GPU support for serverless ML. An ML-based application often consists of multiple stages of work, which form a pipeline or a directed acyclic graph (DAG)~\cite{kannan2019grandslam}. A chatbot, for example, responds to a user's input by going through stages of speech recognition, natural language understanding, speech synthesis, and so on. The SLO of an AI-based application is usually for the {\em end-to-end} latency of the entire process. The schedule and configuration of one stage hence affect other stages, such as how much latency those stages can still have to allow the end-to-end time to meet the SLO, how much resource used in total, whether the data transfer is local or remote, and so on. Neglecting the inter-stage relations to reduce the scheduling complexity is hence not ideal, causing a large loss in quality (36\%-61\% as Section~\ref{sec:detailcomp} shows). (ii) Those that {\em neglect GPU sharing and runtime variations}, represented by Orion~\cite{mahgoub2022orion} and Aquatope~\cite{zhou2023aquatope}. These proposals regard a GPU as a single device attached to a CPU and schedule jobs purely based on CPU and memory availability. Moreover, they are rigid: They pick a good configuration for every stage of the application before its execution, and then stick to the configuration throughout that execution, regardless of whether the actual latencies of some functions differ substantially from the expectation or how the resource availability changes as the execution reaches later stages. The strategy also leads to a large loss in scheduling quality (46\%-80\% as Section~\ref{sec:detailcomp} shows).

Addressing the limitations is essential for unlocking the full potential of GPU for serverless ML. It is, however, not easy: The solution must be efficient enough to handle the enormous configuration space and agile enough to adapt to the dynamic changes of the system.

This paper presents our solution, a new scheduling algorithm named ESG (which stands for \underline{E}fficient Serverless Scheduling for \underline{S}harable \underline{G}PUs), which, for the first time, addresses all those challenges at the same time. ESG treats sharable GPU as a first-order factor in scheduling. For scheduling \underline{efficiency}, ESG employs A*-search and introduces a {\em dual-blade pruning} to dramatically prune the huge search space of schedules without compromising the quality. For \underline{scalability}, ESG further introduces {\em dominator-based SLO distribution} to prevent space explosion. For the \underline{quality} of the scheduling results, ESG takes an {\em optimality-guided adaptive approach}. Rather than deciding the resource assignment of all functions in a workflow at the beginning and keeping it unchanged throughout the DAG execution as previous solutions~\cite{mahgoub2022orion,mahgoub2022wisefuse} do, ESG revisits and adjusts the schedules before the dispatch of every serverless function to adapt to the dynamic changes of the environment and the performance variability of the functions. 

We have evaluated ESG on a series of workloads involving six DNNs and compared the performance with four state-of-the-art serverless scheduling solutions, INFless~\cite{yang2022infless}, FaST-GShare~\cite{gu2023fast}, Orion~\cite{mahgoub2022orion} and Aquatope~\cite{zhou2023aquatope}. The results show that the new scheduling algorithm ESG is effective in addressing the new challenges, consistently achieving the highest SLO hit rates with significantly lower resource usage, exhibiting a notable 61\%-80\% improvement in SLO hit rates and saving 47\%-187\% in costs than INFless and Fast-GShare, especially in challenging scenarios.

Overall this work makes the following major contributions:

\begin{itemize}
\item It provides ESG, the first scheduling algorithm that simultaneously tackles inter-function relations, GPU sharing, batching, and runtime variations.

\item It introduces a set of novel optimizations to ensure high efficiency and scalability of the scheduling algorithm.

\item It empirically confirms the effectiveness of ESG by comparing it with the state of the art represented by four previous scheduling algorithms. 
\end{itemize}

\section{Background}
\label{sec:serverlessbg}


 
\begin{figure}
\centering
\includegraphics[width=0.9\columnwidth]{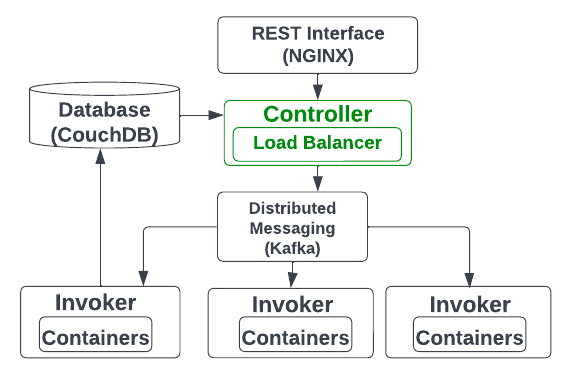}
\caption{OpenWhisk architecture. Controller is where scheduling happens.}
\label{fig:openwhisk}
\end{figure}

\noindent\textbf{Serverless architecture.} We use OpenWhisk~\cite{openwhiskplatform} as an example to explain the common architecture of serverless computing. OpenWhisk is an open-source,
distributed Serverless platform that executes functions
\footnote{“Function” in this paper refers to Serverless Function.}
in response to events at any scale. Figure~\ref{fig:openwhisk} shows the architecture of OpenWhisk~\cite{Openwhisk, shahrad2020serverless}.
It has a RESTful API, through which, users can create actions, invoke actions 
and check the action status.
NGINX translates the command and forwards it to the Controller. Controller is where task scheduling happens. It 
maintains the health and remaining capacity of each Invoker (i.e., a computing node), tracks the warm instances on each Invoker, decides the resource assignments to tasks, and assigns tasks to Invokers. Then the Controller sends the invocation message 
to the selected Invoker via distributed messaging component (Kafka).
The Invoker creates an execution environment (Docker container) 
after receiving the invocation message and manages its runtime 
(including stopping the container). 
OpenWhisk sets a fixed 10-minute keep-alive policy for each instance; the Invoker informs the Controller when it unloads a container~\cite{shahrad2020serverless}.

\noindent\textbf{Scheduling of Serverless Functions.} 
There are two main steps in scheduling serverless functions. 

The first step is resource assignment, which determines the amount of resources assigned to a function. It is based on the memory requirement specified in the configuration, which is converted by some platforms into a number of vCPUs (each vCPU is associated with a fixed amount of memory). The scheduler ultimately determines the amount of resources to be assigned to that instance, which may be more than required to achieve better performance. OpenWisk uses the required amount by default, while recently schedulers~\cite{mahgoub2022orion, akkus2018sand, dukic2020photons, qiu2022simppo, zhou2023aquatope} used a more sophisticated algorithm to decide the assignment of resources. 

The second step is to map a function instance to an Invoker (e.g., a computing node). 
The controller maintains the available CPU resources of each Invoker according to its memory usage. In OpenWhisk, the controller picks one of the Invokers based on the hash value 
calculated by the namespace and action of the function.
The generated index is called "home-invoker"~\cite{openwhiskplatform}, which is the invoker where the future instances of the function will reside by default. But if that invoker becomes unhealthy or lacks capacity, the scheduler will look for other invokers through a deterministic search. The mapping scheme is designed to get more warm starts.

\noindent\textbf{GPU Sharing.}
GPU sharing is to allow multiple processes to share a single GPU for their executions. Modern GPUs support time sharing (one after the other) and spatial sharing. Spatial sharing is essential for ML inferences because an inference by an ML model often uses only a fraction of the massive parallel computing capacity of a GPU. Allowing multiple processes to execute concurrently on a GPU is essential for turning its computing power into throughput. Modern GPUs from NVIDIA offer two mechanisms for spatial sharing, Multi-Process Service (MPS)~\cite{mps} and Multi-Instance GPU (MIG)~\cite{mig}. MIG gives better isolations between partitions. MIG's ability to partition a single GPU into multiple hardware-isolated instances not only maximizes resource utilization but also significantly bolsters security, a critical aspect in today's cloud computing landscape. 
Although the size of a MIG partition is fixed at system booting time, one application or serverless function may use more than one MIG partition by launching multiple GPU kernels concurrently.  
\section{Solution: ESG Scheduling Algorithm}
This section presents ESG, our GPU sharing-aware scheduling algorithm that is designed to respond to the unsolved challenges from inter-function relations and dynamic changes of resources. 

\subsection{Overview}


\begin{figure}
    \centering
    \includegraphics[width=\columnwidth]{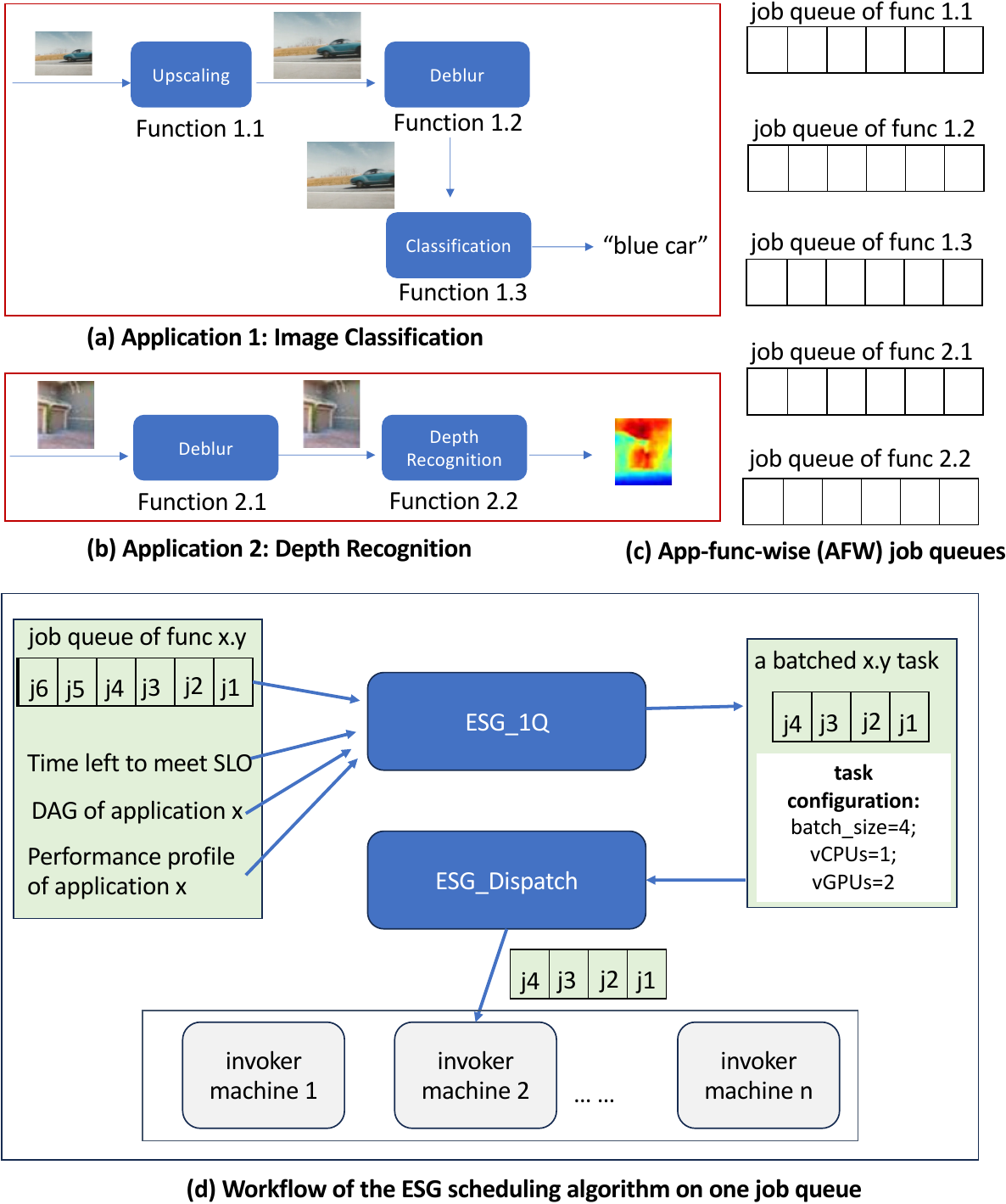}
    \caption{The app-func-wise (AFW) job queues of two example ML-based applications, and the ESG algorithm workflow in handling one job queue.}
    \label{fig:overviewESG}
\end{figure}

ESG runs on the Controller of a serverless platform. Its objective is to maximize the  Service Level Objective (SLO) hits of DNN inferences---that is, make as many DNN inferences deliver their results within the SLO latency as possible---while minimizing resource consumption (and hence cost). The problem is intuitive to understand, so a formal definition is omitted here for the sake of space. Interested readers may see the Appendix~\ref{sec:problemformal} for a formal rigorous definition of the problem.

For easy explanation, Figures~\ref{fig:overviewESG} (a) and (b) illustrate two serverless ML-based applications in the form of DAG. Application 1 uses three DNN serverless functions in a sequence to first upscale an input image, then deblur it, and finally classify it. Application 2 uses two DNN serverless functions to first deblur an input image and then generate a depth image from it. 

ESG introduces {\em application-function-wise (AFW) job queues} to group requests for the same serverless function of the same application together. The five serverless functions in Figure~\ref{fig:overviewESG} (a) and (b) each have their own AFW job queue, as illustrated in Figure~\ref{fig:overviewESG} (c). Notice that even though the same Deblur function is used in both applications, two AFW queues are created, one for each. This design allows ESG to put the functions of the same application on the same machine when possible to reduce the overhead of communications between functions (Section~\ref{sec:esgDispatch}). The AFW queues reside on the Controller. Each of them gets populated as user requests arrive or its predecessor functions produce some triggering outputs. 

\paragraph{Two-step Design.} Figure~\ref{fig:overviewESG} (d) outlines the high-level workflow of the two core algorithms of ESG. The Controller examines the job queues in a round-robin manner. If a job queue is ready to be scheduled, {\em ESG\_1Q}, the first core algorithm of ESG, figures out an appropriate configuration for some jobs in that queue. The configuration includes (i) {\em batch size}: the number of jobs to be scheduled as a group---which we call a {\em task}; (ii) {\em \#vCPUs}: the amount of CPU resources to use; (iii) {\em \#vGPUs}: the amount of GPU resource to use. (Section~\ref{sec:resandtaskmodel} details the resource model).  Then, the second core algorithm of ESG, {\em ESG\_Dispatch}, assigns the task to an appropriate Invoker machine to run. 

Note that ESG\_1Q does not consider current resource availability constraints; as a result, ESG\_Dispatch may not be able to find any Invokers having enough resource to run the task in a configuration found by ESG\_1Q. ESG solves the issue by letting ESG\_1Q output multiple top configurations, forming a {\em configuration priority queue}. ESG\_Dispatch repeatedly dequeus the priority queue until it finds an Invoker that has enough resources to handle the configuration. If none of the configurations works, the scheduler records that AFW job queue in a {\em recheck list}, and moves on to the next AFW job queue. Each time it finishes processing a job queue, it tries again on the job queues in the {\em recheck list}; it may succeed now, as the states of the worker nodes have changed. If a queue stays in the {\em recheck list} too long (e.g., 3 rounds), it will be dispatched with the minimum configuration to ensure progress. 

This two-step design is for two reasons: (i) finding optimal configurations takes some time and the states of the workers may have changed during that time; (ii) finding optimal configurations itself is already complicated, adding the dynamic changing machine status into the search process would further complicate the problem. The two-step design decouples the complexities and hence simplifies the solution. Meanwhile, by postponing the resource availability check to the second step, it is able to consider the current state of workers. 

The two core algorithms both feature some novel designs, which help ESG address the two key challenges in adding sharable GPU into serverless computing, namely, the dramatically expanded resource assignment space and the increased complexity in handling dynamic changes in resource availability and function running times. We next first explain the resource and task models used in this work and then detail both algorithms.

\subsection{Resource Model and Task Model}
\label{sec:resandtaskmodel}

\noindent\textbf{Resource model.}
A change to the resource model for heterogeneous serverless computing is the inclusion of GPU units. Modern GPUs are equipped with built-in mechanisms to support GPU partition and sharing. Multi-instance GPU (MIG) in NVIDIA GPU, for instance, allows GPUs to be securely partitioned into up to seven separate GPU Instances for CUDA applications. With MIG, each instance's processors have separate and isolated paths through the entire memory system, including the on-chip crossbar ports, L2 cache banks, memory controllers, and DRAM address buses that are assigned uniquely to an individual instance. We accordingly integrate the notion of vGPU into the serverless computing resource model. Here, each vGPU is equivalent to the minimum GPU partition of the sharing system (in our case, MIG). We assume each GPU is partitioned into the maximum number of MIG instances (7 for A100). 

For a modern GPU-supported container, when it is launched, it can be set to use one or more vCPUs and one or more vGPUs. When it is set to use multiple vGPUs, it can use them by invoking multiple GPU kernels concurrently, one on each. 

Current serverless platforms typically associate a certain amount of memory with one vCPU, which simplifies resource allocation: the platform can use vCPU as the allocation unit without explicitly allocating memory. GPU memory is associated with vGPU in a similar way, naturally enabled by MIG. We, however, do not associate vGPU with vCPU but make them separate resources for allocation. It is because there is no clear correlation between the amount of CPU usage and the amount of GPU usage in applications. 


\noindent\textbf{Task model.}
The applications targeted in this work are ML inference applications. (ML training is out of the scope.) 
There are two special extensions to the task model in heterogeneous serverless computing. (i) Each serverless function may now contain both CPU and GPU parts. (ii) For ML inferences, even though requests may come one by one, the inference function is often written in a way that it can accept a batch of requests and process them together at one invocation of the function. If the function is given multiple vGPUs, it automatically uses data parallel inferences by launching multiple GPU kernels with each processing one portion of the batch on one vGPU. We call the inference of one request a \texttt{job} and the set of jobs processed by an invocation of a serverless function a \texttt{task}. 

\subsection{ESG\_1Q Algorithm}
\label{sec:esg1q_alg}

\begin{figure*}
    \centering
    \includegraphics[width=0.9\textwidth]{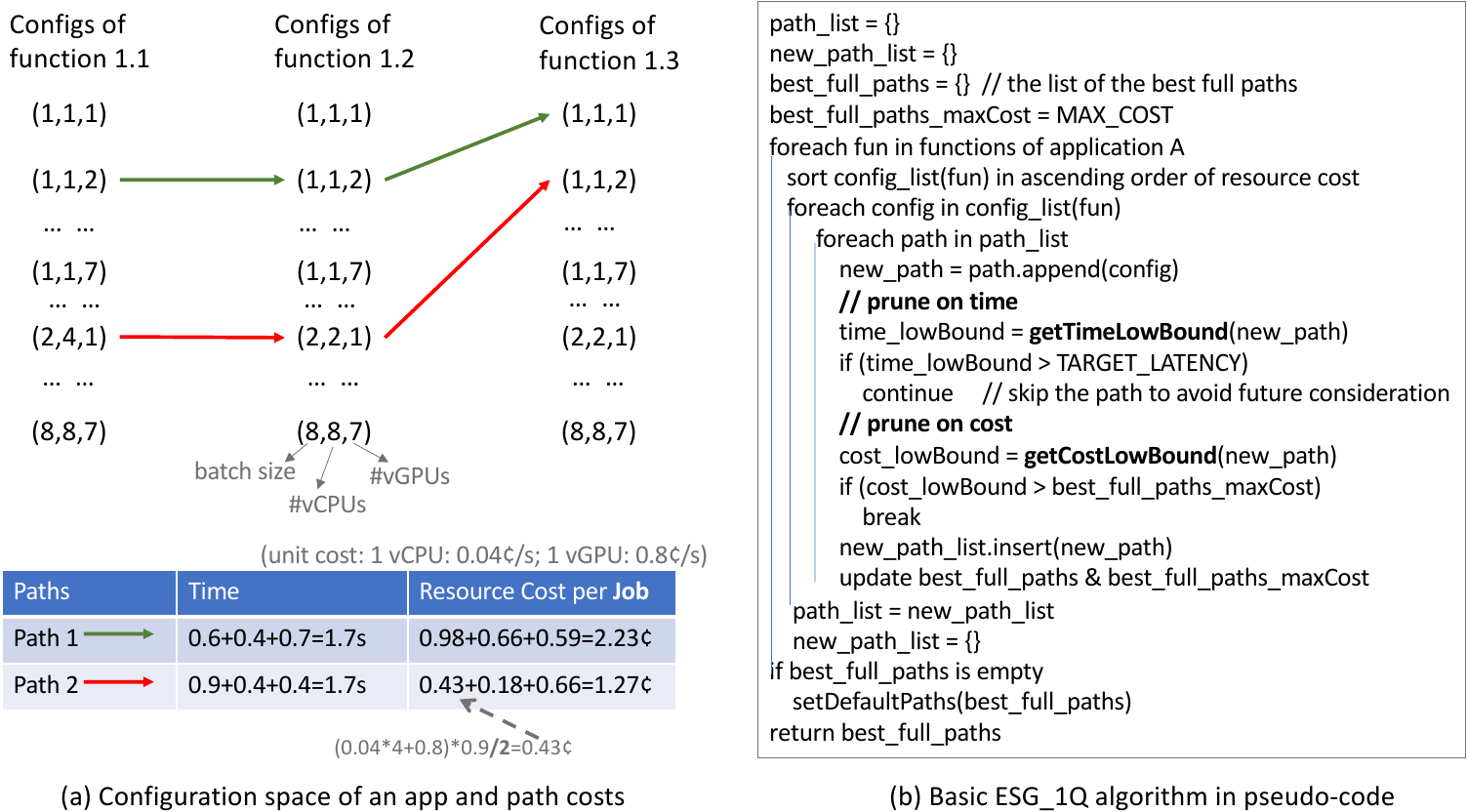}
    \caption{(a) {\bf Top:} Example of the configuration space of a three-function application and two configuration paths in the space. {\bf Bottom:} the time and {\em per-job} resource costs of the two paths. (b) Basic ESG\_1Q algorithm in pseudo-code.}
    \label{fig:esq1q_example}
\end{figure*}


As Figure~\ref{fig:overviewESG}(d) shows, {\em ESG\_1Q} tries to identify a good configuration for the jobs in an AFW queue so that the application can complete within the SLO latency with the minimum resource cost. 

There are two complications: (i) speed-cost tension; a configuration that leads to faster execution tends to incur a higher cost. (ii) Inter-function dependence; the configuration selected for one function determines how much time is left for other functions in the application and hence affects how to best configure the other functions. 

Our approach is to convert the problem into a path-finding problem in the configuration space of an application. The top part of Figure~\ref{fig:esq1q_example} illustrates the configuration space of a 3-function application. A path from the leftmost column to the rightmost column specifies the configurations for each of the three functions of the application. The red path in Figure~\ref{fig:esq1q_example}, for instance, corresponds to a case where the first function processes two requests each time with four vCPUs and one vGPUs, the second function processes two requests with two vCPUs and one vGPU, and the third processes one request with one vCPU and two vGPUs. 

Different paths result in different times and resource costs as exemplified by the table in Figure~\ref{fig:esq1q_example}(a). The Controller can estimate the times with performance profiles of the functions and calculate the costs based on the unit costs of vCPU and vGPU and the running times. 

With that formulation, the goal of ESG becomes finding the path in the configuration space of the application that meets the SLO latency and has the lowest resource cost. So each time, what ESG\_1Q returns is not just a configuration good for the current function, but a sequence of configurations good for the whole application. It is important to note that unlike previous methods, Orion~\cite{mahgoub2022orion} and Aquatope~\cite{zhou2023aquatope}, ESG calls ESG\_1Q again when later functions in the application are to be scheduled, so that it can accommodate the dynamic resource changes and running time fluctuations. 


\paragraph{A*-Search with Dual-Bladed Pruning in ESG\_1Q}
ESG\_1Q uses A*-search as the basis for the path finding, and proposes two techniques, {\em Dual-bladed pruning} and {\em dominator-based SLO distribution}, to ensure the efficiency and scalability of the algorithm for serverless scheduling. 

The A* search algorithm is an efficient and popular path finding and graph traversal method~\cite{hart1968formal}. It efficiently finds the shortest path from a start node to a target node in a weighted graph. A* uses a heuristic to estimate the cost to reach the goal from each node, guiding the search towards the target more efficiently than algorithms like Dijkstra's, which only consider the actual cost from the start node. The total cost f(n) of a node n is calculated as f(n) = g(n) + h(n), where g(n) is the cost from the start node to n, and h(n) is the estimated cost (heuristic) from n to the goal. A* is a best-first search algorithm, meaning it explores a path that appears to be most promising by using the cost function f(n). It prioritizes paths that are expected to lead more quickly to the target. The algorithm maintains a priority queue (often implemented as a min-heap) of nodes to be explored, sorted by their f(n) value. A* is both \underline{complete and optimal}, meaning it will always find a solution if one exists, and the solution will be the shortest possible path, provided that the heuristic function h(n) is admissible (it never overestimates the true cost) and consistent (monotonic).

ESG\_1Q builds on A*, the optimality of which makes ESG scheduling optimality-guided. ESG\_1Q meanwhile improves the path finding with {\em dual-bladed pruning}. Figure~\ref{fig:esq1q_example} (b) outlines the basic idea. 
It prunes the search space based on both running time and resource usage, efficiently avoiding exploration of unnecessary subspaces. The pruning effectively estimates the lower and upper bounds of the costs of a path. When a partial path $p$ forms, ESG\_1Q calculates three bounds, $tLow$, $rscLow$, and $rscFastest$, and uses them for pruning, as follows:

\begin{itemize}
\item $tLow$: the lower bound of the time cost of all paths prefixed by $p$. ESG\_1Q estimates it by summing the time of the functions in $p$ (obtained from the profiles) and the minimum time of each function (among all its possible configurations) not covered by $p$. It is used in time-based pruning (function getTimeLowBound in Figure~\ref{fig:esq1q_example} (b)). 
\item $rscLow$: the lower bound of the resource usage of all paths prefixed by $p$. ESG\_1Q estimates it by adding the resource usage of the functions in $p$ (obtained from the profiles) and the minimum resource usage of each function (among all its possible configurations) not covered by $p$. It is used in the cost-based pruning (function getCostLowBound in Figure~\ref{fig:esq1q_example} (b)). 
\item $rscFastest$: the summation of the resource consumed by $p$ and the resource consumed by other functions when they run the fastest. It is used in updating \\best\_full\_paths\_maxCost in Figure~\ref{fig:esq1q_example} (b) to tighten the bound for cost-based pruning. 
\end{itemize}

Note that for ease of understanding, Figure~\ref{fig:esq1q_example} (b) shows only the basic design of ESG\_1Q; Omitted details (e.g., the use of priority list for best-first search) are documented in Appendix~\ref{sec:pseudocode}.

\paragraph{Dominator-based SLO Distribution for Scalability}

Even with Dual-blade pruning, when dealing with lengthy call sequences within an application, the algorithm's execution time can still be exceedingly long. To address this issue, we introduce a strategy named {\em dominator-based SLO distribution}. It uses {\em stage grouping} to divide the functions of the application into several function groups, uses  {\em SLO distribution} to assign a specific SLO latency to each group, and then applies the ESG\_1Q search algorithm to each individual group. To ensure maximal quality, the groups are maximized under the constraints of tolerable schedule latency.  
This strategy provides a way to strike a good balance between the scheduling scalability and the benefits of considering inter-function relations in scheduling. 

For this approach to work, we need to determine how to group functions and assign individual SLOs to these groups. This solution should be adaptable to both linear pipelines and more complex DAGs with splits and joins.

\begin{figure*}
    \centering
    \includegraphics[width=0.8\textwidth]{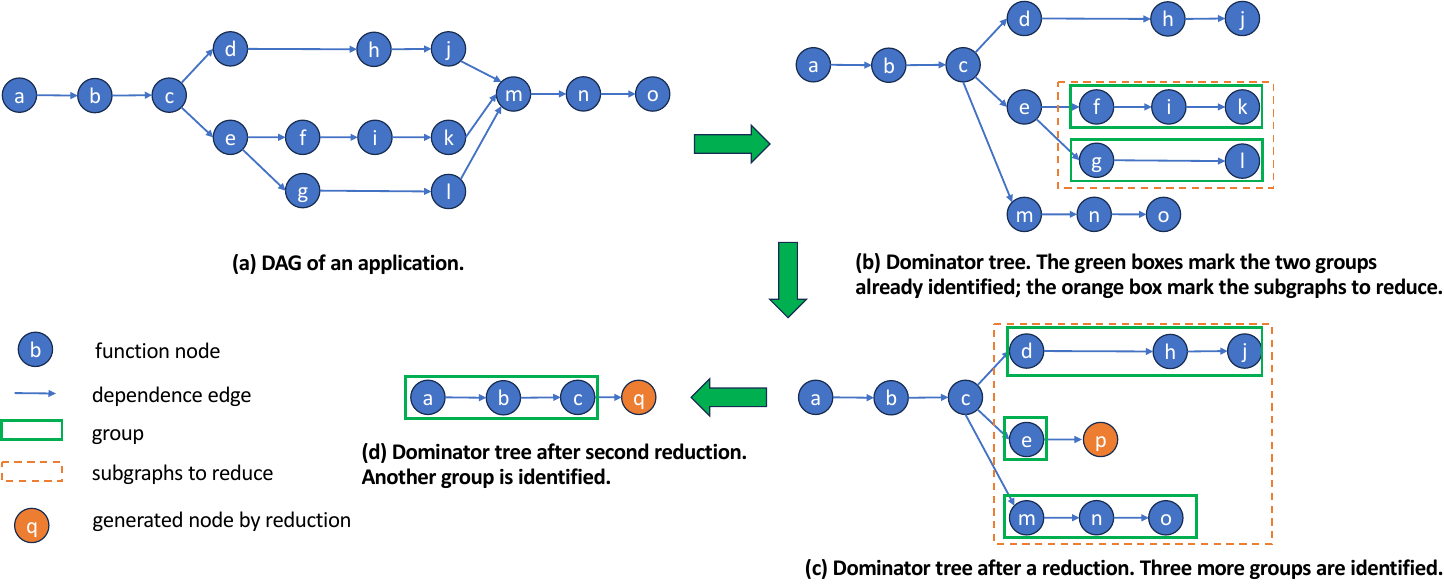}
    \caption{Illustration of dominator-based SLO distribution.}
    \label{fig:reduce}
\end{figure*}

Our solution is a {\em reduction-based hierarchical} method. It is based on an observation that the DAGs in serverless applications are usually hierarchically reducible, as Figure~\ref{fig:reduce} illustrates. The design is inspired by dominator-based code analysis in compilers~\cite{dragonBook}. The algorithm includes four steps. 

First, it creates the dominator tree of the DAG following traditional compiler-based code analysis~\cite{dragonBook}. In a dominator tree, each edge indicates an immediate domination relation in the DAG\footnote{A dominates B if all paths from the root to B must first reach A; an immediate dominator is the closest dominator except the node itself~\cite{dragonBook}.}, as illustrated in Figure~\ref{fig:reduce}. The data structure offers the basis for an ordered traversal and reduction in processing.

Second, it labels each node in the dominator tree with the {\em average normalized length (ANL)} of its function (say $f_i$), $average_c(l_{f_i}(c))$, where $c$ is a configuration, $l_{f_i}(c)$ is the normalized length of the function $f_i$ in configuration $c$, calculated as $t_{f_i}(c)/\sum_j t_{f_j} (c)$; $t_{f_j}(c)$ is the execution time of function $f_i$ in configuration $c$, read from the performance profile.

Third, it traverses the dominator tree in a post-order (children before parents). At each node (say $x$), if it has more than one child, it calls subroutine {\em reduce(x)} to first reduce and reorganize its descendants. The {\em reduce} operation is to combine its branches into one node, as illustrated in Figure~\ref{fig:reduce} (c). The ANL of that new node is the maximum of the sums of the ANL of all the branches. After that, the algorithm checks if the parent of $x$ has more than one child and if so, it calls the subroutine {\em slo\_group(x)} to partition $x$ and its descendants into one or more groups.   
Note that thanks to {\em reduce}, the descendants of $x$ are guaranteed to form a single list. The partition simply groups consecutive $g$ nodes into one group, where $g$ is the maximum group size (Section~\ref{sec:sensitive} reports how $g$ affects performance). An exception is the nodes generated by reduction, each of which stays as an individual node to prevent their subsumed groups' sizes from being bloated. This process continues until the algorithm reaches the root. Throughout the process, the algorithm records the reduction process for the next step to use. 

Finally, the algorithm assigns SLO latency to each of the groups. This process reverses the reduction process. It starts from the final form of the reduced dominator tree (which is a list) and partitions the end-to-end SLO latency proportionally based on the ANLs of its groups. It then calls subroutine {\em slo\_assign(x)} for each reduced node in the current list to partition the SLO latency of $x$ and assigns the partitions to the nodes that $x$ subsumes. This process continues until every group gets its SLO quota. 




\subsection{ESG\_Dispatch: Mapping to Worker Nodes}
\label{sec:esgDispatch}

After ESG\_1Q, \textbf{ESG\_Dispatch} maps the current group of jobs to an Invoker node. As we introduced
in Section~\ref{sec:serverlessbg}, the OpenWhisk scheduling always chooses the home-invoker first; if not feasible, it tries other worker nodes.
Our algorithm chooses the home-invoker for the first function in the workflow. For other functions, it would try to choose the invoker that runs its predecessor function in the workflow. This locality-sensitive treatment is possible thanks to our introduced AFW queues. It helps reduce data transfer, as communications on the same node can use local file systems rather than remote storage. This consistent policy is beneficial for getting warm starts. 
If the home-invoker or predecessor-invoker does not have enough available resources,
the algorithm will try other warm Invokers. 
If it fails, it will check other cold invokers and choose the one with the most available resources.

\section{Methodology for Evaluation}
\label{sec:methodology}

To evaluate the efficacy of the scheduling algorithm, we create a framework that can emulate various serverless workloads and scenarios. The emulations are based on actual performance of the serverless functions measured on actual machines in various configurations (batch size, CPU and GPU resource allocations). The machine is as specified in Table~\ref{tab:hardwareinfo}. To accommodate the impact of other runtime factors on the performance, the emulations add Gaussian noises to the performance. The emulation is equipped with a workload generator, which generates workloads by sampling the set of serverless functions randomly based on a specified arrival rate. The set of workloads we considered in the evaluation are further detailed in Sec~\ref{sec:application}. The hardware resource of the considered testbed in the emulations consists of 16 nodes, with each equipped with 16 vCPUs and 1 GPU (up to 7 vGPUs by MIG). The scheduler and job dispatching implementation is based on the controller in OpenWhisk~\cite{Openwhisk}. 




We use proxy threads to monitor the function call intervals, predict subsequent invocations, and preemptively warm up instances.
Studies have shown that pre-warming can help reduce delays caused by cold starts. There have been several methods proposed before, with some using ML models~\cite{zhou2023aquatope} and others~\cite{mahgoub2022orion, mahgoub2021sonic, yang2022infless, shahrad2020serverless} using a histogram-based policy to adjust container keep-alive times. We use a lightweight method for prewarming. It uses Exponential Weighted Moving Average (EWMA)~\cite{ewma} to predict the invocation intervals of functions and pre-warms the function instances accordingly. After pre-warming, ESG uses the same keep-alive policy as OpenWhisk, to keep the instance alive for 10 minutes.

\begin{table}[!h]
\centering
\scriptsize
\caption{Experimental testbed configuration}
\begin{tabular}{|c|c|}
\hline
{\textbf{CPU device}}    & \begin{tabular}[c]{@{}c@{}}AMD EPYC 7302P 16-Core Processor\end{tabular} \\ \hline
                                                                                                {\textbf{CPU Mhz}}       & 1499.866                                                                    \\ \hline 
                                                                                                {\textbf{CPU memory}}    &       \begin{tabular}[c]{@{}c@{}}128GB DDR4 3200MHz ECC DRAM\end{tabular}                                                         \\ \hline 
                                                                                               {\textbf{GPU device}}    & NVIDIA A100 80GB                                                            \\ \hline 
                                                                                               {\textbf{GPU memory}}    & 80GB                                                                        \\ \hline
                                                                                               {\textbf{MIG instances}} & Up to 7 MIGs@10GB                                                                                                                                \\ \hline
\end{tabular}
\label{tab:hardwareinfo}
\end{table}

\subsection{Applications}\label{sec:application}
 The workloads in the experiments are series of calls to four applications, with each consisting of a sequence of DNN inferences. Table~\ref{tab:functions} reports the source, inputs, DNN models, execution time in the minimum configuration (1vCPU, 1vGPU, batch size=1), and cold start time of the functions. The four applications are detailed as follows:
 
\begin{table}[!t]
\centering
\scriptsize
\tabcolsep=0.01cm
\caption{Serverless Functions}
\begin{tabular}{|c|c|c|c|c|}
\hline
  \textbf{Function name}                & \textbf{\begin{tabular}[c]{@{}c@{}}Execution \\Time (ms)\end{tabular}} & \textbf{\begin{tabular}[c]{@{}c@{}}Cold start \\time (ms)\end{tabular}} & \textbf{\begin{tabular}[c]{@{}c@{}}Input image \\size (MB)\end{tabular}} & \textbf{Model}               \\ \hline
\begin{tabular}[c]{@{}c@{}}Super resolution~\cite{ledig2017photo}\end{tabular}  & 86                                                            & 3503                                                           & 2.7                                                             & SRGAN               \\ \hline
\begin{tabular}[c]{@{}c@{}}Segmentation~\cite{chen2017rethinking,deeplabv3}\end{tabular}      & 293                                                           & 16510                                                          & 2.5                                                             & \begin{tabular}[c]{@{}c@{}}deeplabv3\_resnet50\end{tabular} \\ \hline
\begin{tabular}[c]{@{}c@{}}Deblur~\cite{deblur}\end{tabular}           & 319                                                           & 22343                                                          & 1.1                                                             & DeblurGAN            \\ \hline
\begin{tabular}[c]{@{}c@{}}Classification~\cite{he2016deep,resnet}\end{tabular}    & 147                                                           & 18299                                                          & 0.147                                                           & ResNet50    \\ \hline
\begin{tabular}[c]{@{}c@{}}Background removal~\cite{Qin_2020_PR}\end{tabular} & 1047                                                          & 3729                                                           & 2.5                                                             & $U^{2}$ Net \\ \hline
\begin{tabular}[c]{@{}c@{}}Depth recognition~\cite{ranftl2020robust}\end{tabular} & 828                                                           & 16479                                                          & 0.648                                                           & MiDaS               \\ \hline
\end{tabular}
\label{tab:functions}
\end{table}

\begin{itemize}
    \item \textbf{Image classification:} It detects and classifies objects, important for autonomous driving~\cite{fujiyoshi2019deep} and other domains. Its workflow is to use super-resolution~\cite{ledig2017photo} first to enhance the clarity of an image, and then use segmentation~\cite{chen2017rethinking,deeplabv3} followed by classification~\cite{he2016deep,resnet} to identify the objects.  
    \item \textbf{Depth recognition:} This application measures the distance of an object from a camera, which is essential to 3D scene reconstruction and augmented reality~\cite{kumar2020lidar}. Its workflow uses deblur~\cite{deblur} followed by super-resolution first to enhance the image and then recognize image depth with another DNN~\cite{ranftl2020robust}.
    \item \textbf{Background elimination:} This application eliminates unnecessary and unwanted items and objects from images~\cite{bau2020semantic}. Its workflow is super-resolution followed by deblur to enhance the image clarity and then uses background removal DNN~\cite{Qin_2020_PR} to eliminate the background.
    \item \textbf{Expanded image classification:} This is a more advanced image classification application with a longer workflow: deblur, followed by superresolution, background removal, segmentation, and classification.
\end{itemize}

\begin{figure}[!h]
    \centering
    \includegraphics[width=0.8\columnwidth]{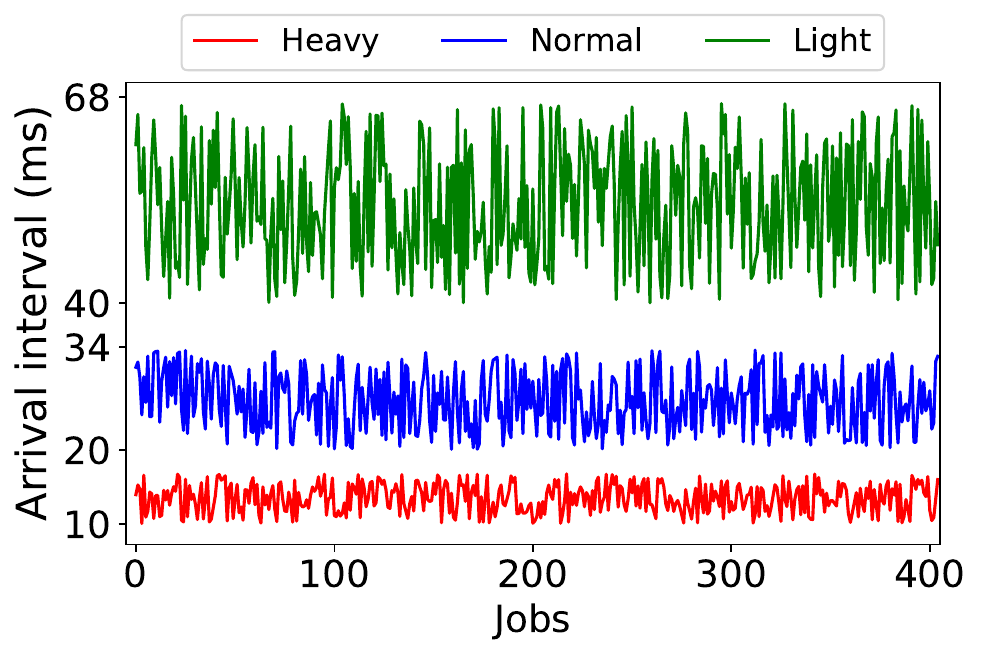}
    \caption{Job arrival intervals used in the evaluation part for different workload settings.}
    \label{fig:workloads}
\end{figure}

For the workload setting, we examined the traces published by Azure~\cite{shahrad2020serverless} to determine job arrival rates. We get the job arrival rates of every minute from the Azure traces, based on which we derived three situations for our DNN applications respectively with {\bf heavy}, {\bf normal}, and {\bf light} workloads. In each workload, one of the four DNN applications is randomly picked to get invoked in each time interval. The length of a job arrival interval is selected randomly in ranges [10--16.8ms], [20--33.6ms], and [40--67.2ms] respectively in the three situations. Figure~\ref{fig:workloads} shows the distribution of the job arrival intervals in the three situations.

We tested three SLOs. Let $L$ be the time needed by the application to complete its entire workflow when it runs alone with the minimum configuration. (i) In the {\bf strict} setting, a SLO hit occurs when the application completes within $0.8\times L$. (ii) In the {\bf moderate} setting, a SLO hit happens when the application completes within $1\times L$. (iii) In the {\bf relaxed} setting, a SLO hit happens when the application completes within $1.2\times L$. 

Those three levels of requirements correspond to users' possible expectations of service: strict for the light case, moderate for the normal case, and relaxed for the heavy case, denoted as {\bf strict-light}, {\bf moderate-normal}, and {\bf relaxed-heavy} workloads in the evaluation section. 

Following AWS EC2 pricing~\cite{yang2022infless}, we set the price of a vCPU to 0.034\$/hour. Based on the pricing of an entire GPU on AWS, we divide it by \# of vGPUs and set the price of a vGPU to 0.67\$/hour. 


\subsection{Comparison Counterparts}


We evaluate our ESG algorithm by comparing it with four state-of-the-art scheduling algorithms, including \textbf{INFless}~\cite{yang2022infless} and \textbf{FaST-GShare}~\cite{gu2023fast}, the latest algorithms for sharable GPU-based serverless ML, as well as the \textbf{best-first search} algorithm in Orion~\cite{mahgoub2022orion} and the \textbf{Bayesian Optimization-based scheduling} in Aquatope~\cite{zhou2023aquatope}. The original Orion~\cite{mahgoub2022orion} and Aquatope~\cite{zhou2023aquatope} do not support GPU sharing, but their scheduling algorithms represent the latest search-based scheduling and model-based scheduling, respectively. For comparison, we extend both with GPU sharing support, as explained below.


\noindent\textbf{InFless:} InFless schedules jobs by enumerating the configurations for each function without considering the inter-function relations. In worker node selection, a resource efficiency metric is used to maximize the throughput while reducing resource fragmentation. InFless provides no method for distributing an application's SLO to its functions. Our experiment follows a prior work~\cite{kannan2019grandslam} to do the distribution based on the average service times of the functions. 

\noindent\textbf{FaST-Gshare:} This work uses FaST-Manager to manage spatio-temporal resources for GPU multiplexing. It also employs an enumeration-based scheduling algorithm which enumerates the configurations based on throughput performance metrics. Its node selection tries to minimize GPU resource fragmentation. It offers no method for distributing an application's SLO either. We apply the same method as in INFless.


\noindent\textbf{Aquatope:} Aquatope relies on an offline training process, in which the application of interest is profiled in many sample executions based on Bayesian Optimization (BO), through which it builds up a performance model and learns about the statistically good configurations for every stage in the application (encoded in an {\em acquisition function}). In deployment, it uses the learned best configurations for the application. Specifically, the training process starts with 100 bootstrapping samples, iterates 50 rounds (we sample five configurations in each round), and selects the best configuration. The nature of its reliance on offline training makes it unable to adapt to dynamic workload changes, as shown in the next section. 

\noindent\textbf{Orion:} Orion creates a performance model to address runtime variations, consolidates parallel invocations into a single virtual machine (VM), and implementes instance pre-warming to eliminate cold starts. Its scheduling uses {\em best-first search}, which creates a priority queue, in which all new states are added. Adding vGPU into the algorithm, we expand its state definition to a vector of (batch size, \#vCPUs, and \#vGPUs), one for each stage. The algorithm examines possible states, with each new state increasing the current state in one dimension of the configuration vector, and the start state ${S_0}$ has the minimum values for every stage function. The scheduling method decides the schedule for all the stages of an application at the invocation of the first stage; no dynamic adaptation between stages. 
As in the original work, P95 latency is used as the search goal. The configuration with the closest latency to the SLO is returned when the search exceeds a cut-off time (e.g., 100ms) before reaching the goal.




The GPU-sharing and batching policy can improve resource utilization, as shown in Section~\ref{sec:batchandsharing}.
So, to evaluate the effectiveness of the scheduling algorithm, in our comparisons, we enable the same GPU-sharing and batching for all the scheduling algorithms; the same data locality and pre-warming policy proposed in this work are also used; the only difference is the scheduling algorithm. INFless and FaST-GShare do not follow the data locality policy but their resource fragmentation minimization policy.


\section{Evaluation}
\label{sec:evaluation}

\begin{figure*}
    \centering
    \includegraphics[width=1\textwidth]{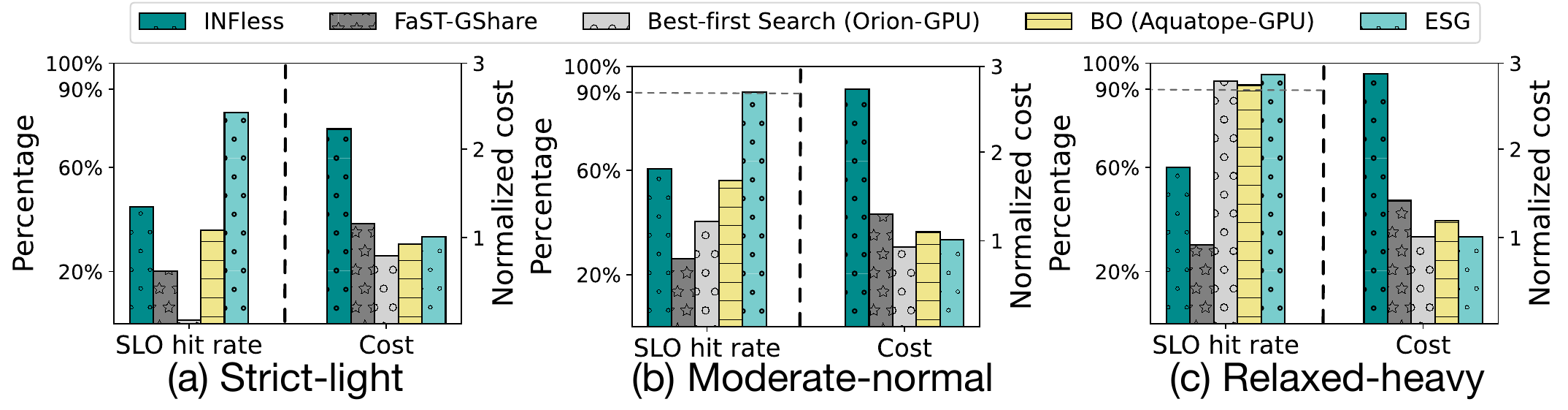}
    \caption{The average SLO hit rate and the cost (normalized to ESG cost) under different SLO and workload settings. The left y-axis is for the SLO hit rate and the higher is better. The right y-axis is for the cost and the lower is better.}
    \label{fig:schedulingCom}
\end{figure*}

\begin{figure*}
    \centering
    \includegraphics[width=1\textwidth]{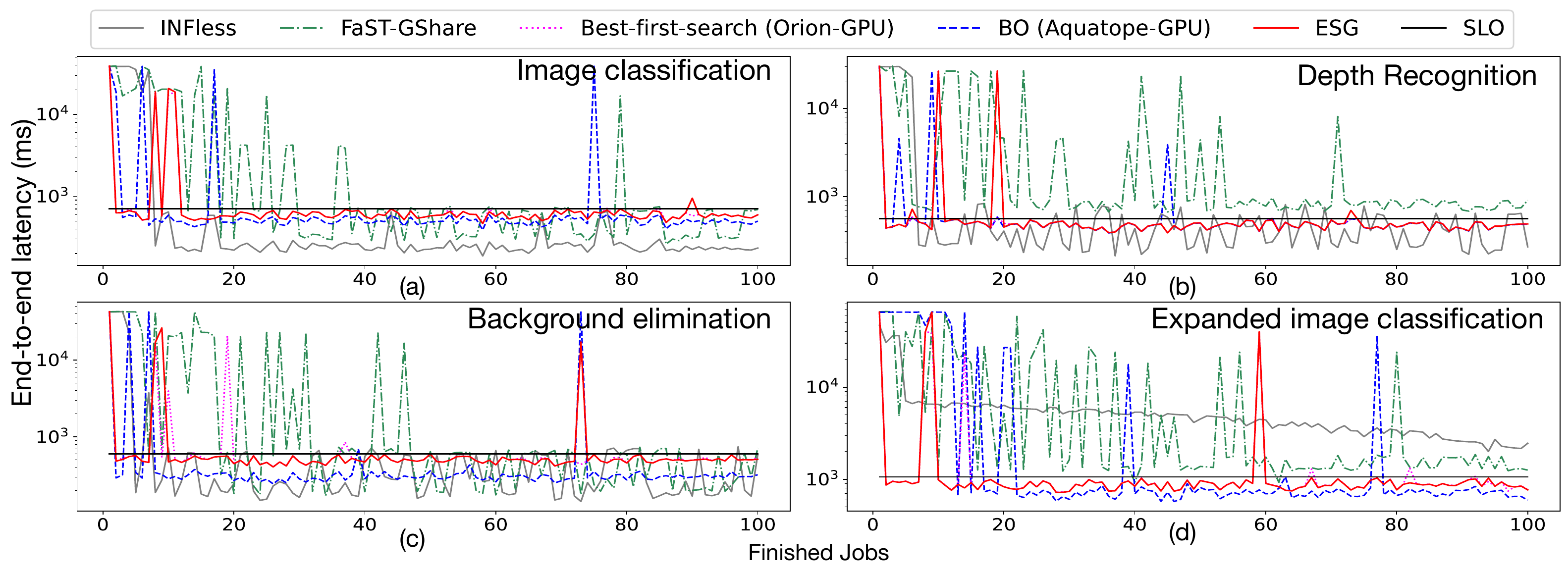}
    \caption{End-to-end latency for each application in relaxed-heavy setting.}
    \label{fig:endtoend}
\end{figure*}

\begin{figure*}[!h]
    \includegraphics[width=.82\textwidth]{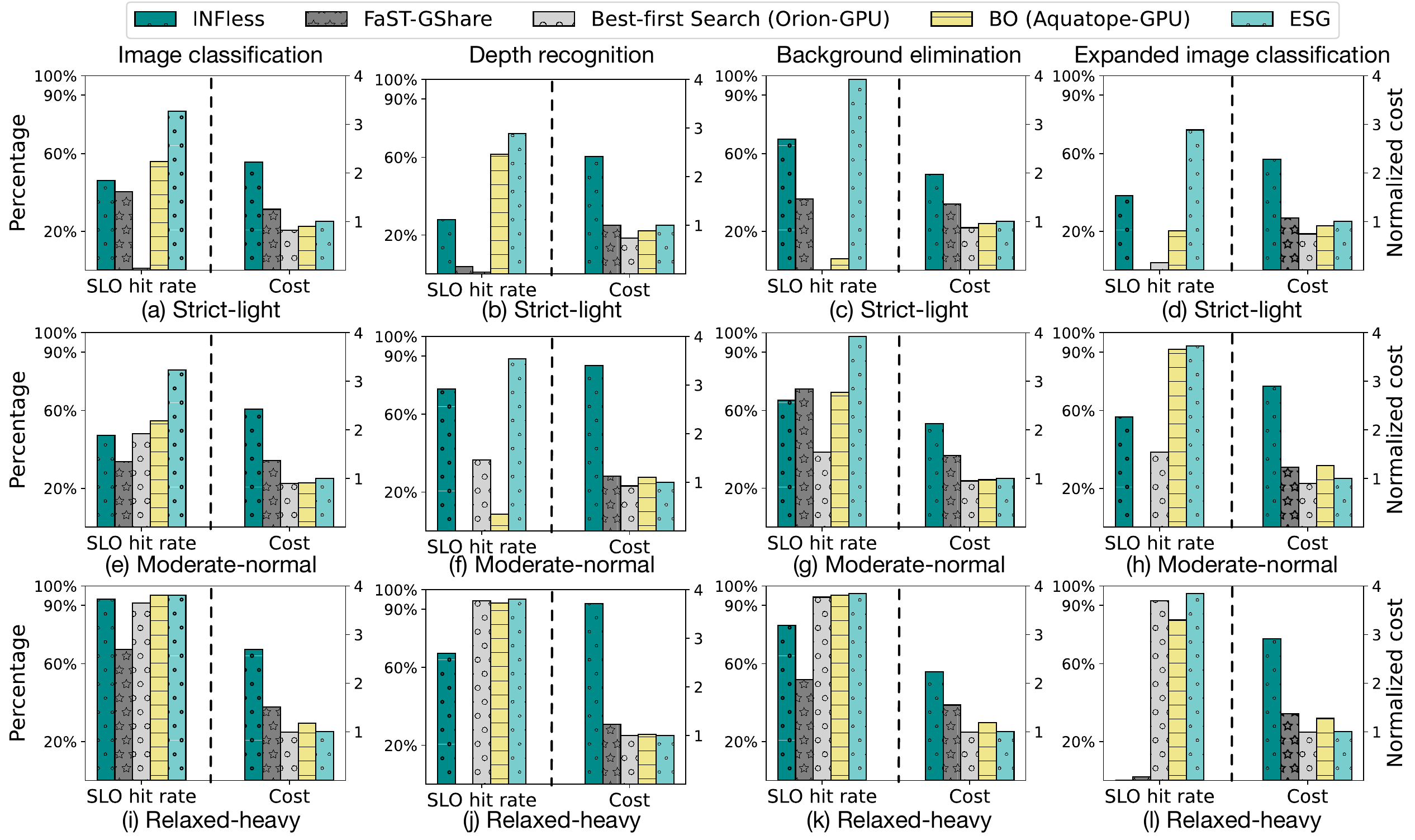}
    \caption{SLO hit rates and cost for each application in three different workload settings.}
    \label{fig:perapp}
\end{figure*}

Our evaluation examines (i) the overall effectiveness of ESG in maximizing SLO hit rates while minimizing the cost; (ii) the reasons for the benefits of ESG over the state-of-the-art scheduling algorithms (INFless~\cite{yang2022infless}, FaST-GShare~\cite{gu2023fast}, Orion~\cite{mahgoub2022orion} and Aquatope~\cite{zhou2023aquatope}); (iii) the overhead and sensitivity analysis of ESG.






\subsection{End-to-End Performance}
\label{sec:detailcomp}

Figure~\ref{fig:schedulingCom} shows the average SLO hit rates and total normalized cost (ESG is 1) for all applications across three situations.
In all three scenarios, ESG consistently shows a high SLO hit rate. Its benefit is especially pronounced in the strict-light scenario: Its hit rate is 46\%-80\% higher than BO and Orion, and 36\%-61\% higher than INFless and Fast-GShare. 
It is noteworthy that ESG achieves the significantly higher SLO hit rates with similar or much less resource cost compared to other methods, as Figure~\ref{fig:schedulingCom} shows.  Detailed SLO hit rates
and cost of each application are shown in Figure~\ref{fig:perapp}. ESG consistently achieves the highest SLO hit rate at a lower cost, whereas INFless consumes the most resources.




For a more detailed view, Figure~\ref{fig:endtoend} shows the end-to-end latencies of each of the four applications in the relaxed-heavy setting.
ESG consistently achieves latencies below but close to the SLO latency. The configurations found by other methods cause the jobs to either run too slow (e.g., FaST-GShare) or to use more resources (e.g., INFless) than necessary at the expense of larger cost or poor performance of other applications. 

For instance, INFless selects configurations that yield lower latencies for applications such as ``image classification'', ``depth recognition'', and ``background elimination'' compared to other methods. This is due to its resource efficiency metric in scheduling, which reduces resource fragmentation and increases system throughput, thereby preferring to utilize all remaining resources in one invoker. However, this approach of allocating excessive resources, as confirmed by the highest resource costs shown in Figure~\ref{fig:perapp}, leads to prolonged waiting times and increased latency for the ``expanded image classification'' application. This application is particularly affected due to its extensive pipeline, which involves more pipeline stages awaiting resources.



\subsection{Detailed Analysis}

In this section, we provide a detailed analysis to examine the reasons for the benefits of ESG over other methods. 


\textbf{Compared to INFless and FaST-GShare:} INFless and FaST-GShare distribute the end-to-end SLO to different stages based on their average service times, without considering their inter-dependencies. Consequently, if certain early stages experience delays due to data transfer overhead and cold start latency, the later stages do not adjust their SLO settings, resulting in prolonged execution time, especially when the application has multiple functions. As demonstrated in Figure~\ref{fig:endtoend}(d), FaST-GShare and INFless always yield the largest latency. Furthermore, in selecting worker nodes, they prefer reducing resource fragmentation rather than focusing on data locality and data transfer latency. This results in even lower SLO hit rates, a fact that is evidenced by the numerous strikes seen in the FaST-GShare curve of Figure~\ref{fig:endtoend}.


\begin{figure}
    \centering
    \includegraphics[width=0.6\columnwidth]{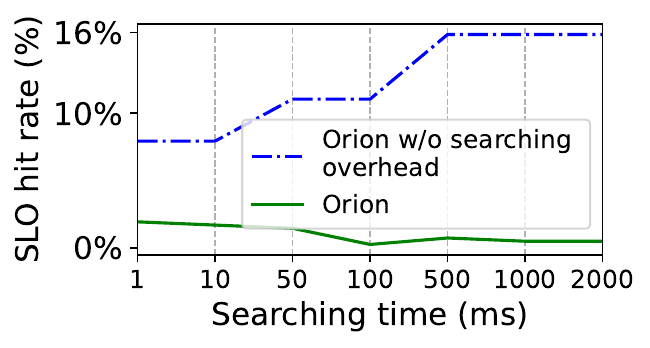}
    \caption{The effect of search time of Orion on the SLO hit rates (strict-light setting).} 
\label{fig:orionsloandcost}
\end{figure}

\textbf{Compared to Orion:} Orion is a search-based method. It faces a trade-off between the search time and the quality of the search result. Figure~\ref{fig:orionsloandcost} shows the tradeoff in the strict-light setting. In this setting, Orion can find quite good configurations. The blue curve in Figure~\ref{fig:orionsloandcost} shows the hit rates of those configurations when search overhead is not counted in. But when the search time is counted in, the 
hit rates drop dramatically, as the green curve in Figure~\ref{fig:orionsloandcost} shows. Moreover, because Orion decides the configurations for all functions in an application when scheduling the first function and does not adjust the configurations of later stages, the configurations are often low in quality or do not even apply. It is especially common when the resource availability changes significantly along time. Table~\ref{tab:missrate} shows the percentage of times when the configurations fail to apply to a function because the batch size in the configuration is even greater than the number of jobs in the queue of that function when it is time to be scheduled. In the moderate-normal and relaxed-heavy settings, because resource availability changes substantially, the percentage is as much as 27--52\%.

ESG overperforms Orion because (i) ESG finds better configurations much faster, thanks to its dual-bladed pruning and dominator-based SLO distribution (overhead analysis shown in the next subsection); (ii) ESG adapts the configurations for every function in a workflow. As a result, with a smaller overhead, ESG produces configurations that meet the SLO and demand fewer computing resources. The lower computing resource demand gives multi-fold benefits. It not only lowers the cost, but also makes the function more likely to fit into the available resource of the predecessor worker node, which in turn leads to better data locality, less communication overhead, and fewer cold starts, which all contribute to the higher SLO hits. 


   \textbf{Compared to Aquatope:} Aquatope relies on a statistical model trained with offline traces. It has negligible scheduling overhead. However, ESG surpasses Aquatope because of the better quality of the configurations given by ESG. As shown in Figure~\ref{fig:workloads}, real workloads and resource conditions continually change. Being a method based on an offline training model, the BO method cannot adapt to dynamic changes. It assumes that configurations remain unchanged unless functions or inputs are modified, but this static scheduling configuration proves inadequate in an ever-changing real-world environment where future job and worker node statuses are uncertain. It is confirmed by the rightmost column in Table~\ref{tab:missrate}, which shows that 59--86\% of the configurations preset by the BO method do not apply in actual executions because the actual queue length is smaller than the batch size in the configuration.

\begin{table}[]
\centering
\scriptsize
\caption{Pre-planned scheduling miss rate}
\begin{tabular}{|c|cc|}
\hline
\multirow{2}{*}{\textbf{System setting}} & \multicolumn{2}{c|}{\textbf{Configuration miss rate}}         \\ \cline{2-3} 
                                         & \multicolumn{1}{c|}{\textbf{Best-first search (Orion)}} & \textbf{BO (Aquatope)} \\ \hline
\textbf{Strict-light}                    & \multicolumn{1}{c|}{9.6\%}                      & 85.5\%      \\ \hline
\textbf{Moderate-normal}                 & \multicolumn{1}{c|}{27.32\%}                    & 59.85\%     \\ \hline
\textbf{Relaxed-heavy}                   & \multicolumn{1}{c|}{51.68\%}                    & 58.72       \\ \hline
\end{tabular}
\label{tab:missrate}
\end{table}

\begin{figure}
    \centering
    \includegraphics[width=0.6\columnwidth]{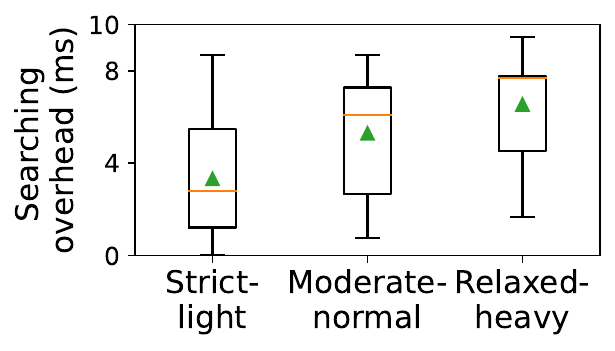}
    \caption{Scheduling overhead distribution of ESG (function group size is 3).} 
\label{fig:searchoverhead}
\end{figure}

\subsection{Overhead Analysis}
\label{sec:overheadanalysis}


Figure~\ref{fig:searchoverhead} reports the scheduling overhead distribution of ESG in the three settings (using the default function group size 3), with the green triangle indicating the average. The searching overhead increases with more relaxed SLO settings. It is because in relaxed SLO settings, more configurations can satisfy the SLO, resulting in fewer configurations being pruned during the search process. But overall, the search overhead is less than 10ms. In comparison, the time taken by a brute-force search would be orders of magnitude higher. In the case where each function has 256 configurations, the search time is 7258ms.



\subsection{Sensitive study}
\label{sec:sensitive}

\begin{figure}
    \centering
    \includegraphics[width=1\columnwidth]{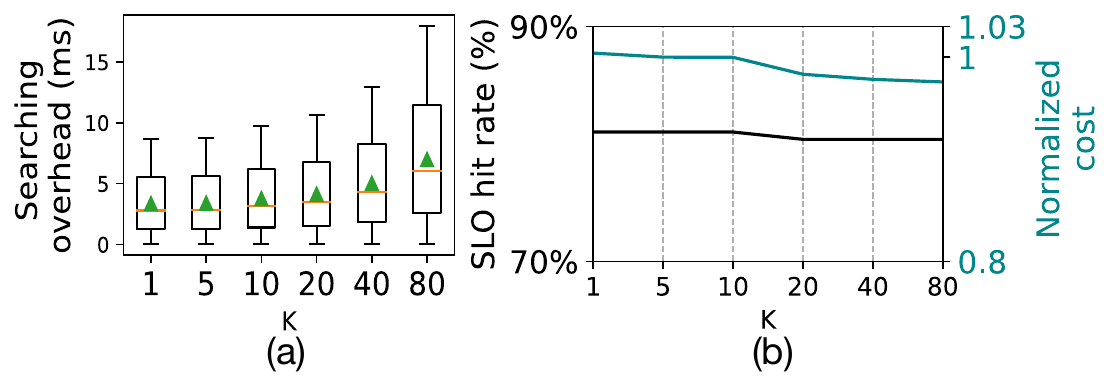}
    \caption{Sensitive study of ($K$) in strict-light setting, the cost of K=5 is set to be 1.} 
\label{fig:K}
\end{figure}

In ESG, there are two parameters, the maximal size of a function group and the number of solutions in the configuration priority queue ($K$). The default maximal group size is set to 3 because when the size increases to 4, the search time jumps to 1201ms (for 256 configurations per function) due to the exponential growth of the configuration space. Regarding the impact of $K$ value, Figure~\ref{fig:K} reports our observations. As $K$ increases from 1 to 80, the average search overhead increases from 3ms to 8ms, the latency remains similar, and the cost decreases slightly. The default $K$ is set to 5 in ESG. 


\subsection{Impact of GPU-sharing and Batching}
\label{sec:batchandsharing}

\begin{figure}[!h]
    \centering
    \includegraphics[width=0.8\columnwidth]{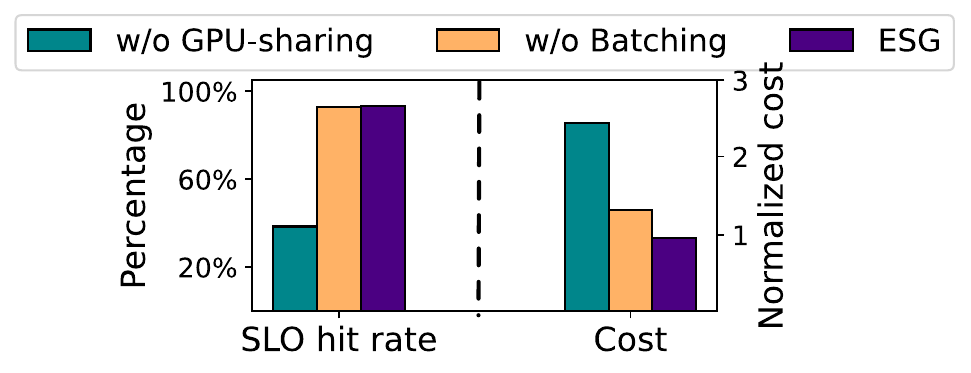}
    \caption{The ablation study in relaxed-heavy setting.}
\label{fig:ablation}
\end{figure}

Through our ablation study, we assess how the GPU-sharing and batching strategies enhance GPU resource efficiency. We individually removed either the GPU-sharing or batching strategy from ESG and contrasted the results with the original ESG. We set a heavy workload in this experiment specifically to underline the effects of the batching strategy. The results indicate that both strategies boost the usage of GPU resources, as evident in Figure~\ref{fig:ablation}.

Without the GPU-sharing strategy, the waiting time is substantially prolonged compared to ESG. This is because jobs are queued, waiting for a GPU (currently in use) to free up. Consequently, the data-locality strategy may falter, leading to substantial data transfer costs and worse SLO hit rates. The batching strategy is crucial in conserving cost, as shown in Figure~\ref{fig:ablation}. The batching policy will delay the execution time; thus, no-batching policy will not decrease the SLO hit rates.

\section{Related Work}


\noindent\textbf{Heterogeneous serverless computing:} Recent studies on GPU-based serverless computing focused on CPU-GPU data transfer~\cite{zhao2023gpu, hong2022gpu} and cold start overhead~\cite{san2023flashpoint, san2023reducing}. Some prior works ~\cite{pemberton2022kernel, lannurien2023herofake} studied the scheduling tasks on GPUs but regarded an entire GPU as the minimal computing unit. Other works~\cite{yang2022infless, gu2023fast, cho2022sla} proposed the GPU-sharing scheduling for inference applications but neglected the inter-function relations of the DAG workflows. There have been efforts to extend serverless computing support to heterogeneous hardware, including FPGAs and Nvidia DPUs~\cite{du2022serverless, vandebon2021scheduling}. Additionally, Dgsf~\cite{fingler2022dgsf} proposed the disaggregated GPU for serverless computing that is orthogonal to our work.


\noindent\textbf{Serverless resource management:} Several works~\cite{yang2022infless, mahgoub2022orion,zhou2023aquatope, qiu2022simppo, kaffes2022hermod, song2023sponge, chen2023olpart} have studied the resource management problem in serverless computing. Orion~\cite{mahgoub2022orion} and INFless~\cite{yang2022infless} are search-based scheduling algorithm. SIMPPO~\cite{qiu2022simppo} used reinforcement learning for serverless resource management. Aquatope~\cite{zhou2023aquatope} is a Bayesian Optimization based scheduling algorithm, which builds upon IceBreaker~\cite{roy2022icebreaker} and CLITE~\cite{patel2020clite} and extends BO in new ways than what was previously done in other BO-inspired solutions like SATORI~\cite{roy2021satori} or Ribbon~\cite{li2021ribbon}, and inspired the OLPart~\cite{chen2023olpart}. 

\noindent\textbf{Optimization of data transfer:} Researchers realize the data transfer between workers and remote database lead to unnecessary latency. FaaSFlow~\cite{li2022faasflow} proposed to partition the workflow into sub-graphs and schedule these functions in one invoker to avoid data transfer. Nightcore~\cite{jia2021nightcore} proposed the internal function calls and low-latency message channels, and efficient threading for I/O to reduce the data transfer latency. Sonic~\cite{mahgoub2021sonic} proposed the hybrid data passing methods, which are direct-message passing and remote storage to reduce the data transfer latency. Palette~\cite{abdi2023palette} used the "colors" to place successive
invocations related to each other on the same executing node.
As shown in previous sections, due to the lack of effective methods to handle large schedule space, prior work left much potential for sharable GPU locked for ML on serverless platforms.

\section{Conclusion}

This study has proposed and evaluated a new algorithm ESG to effectively schedule ML workloads on serverless platforms with sharable GPUs. The search algorithm employs an {\em optimality-guided adaptive} method by combining A*-search and a novel {\em dual-blade pruning} to effectively prune the scheduling space without compromising the quality. Its {\em dominator-based SLO distribution} offers a way to keep the algorithm scalable. Across a diverse set of real-world serverless applications, ESG gives the highest SLO hit rates, while significantly reducing the cost.

\begin{acks}
This material is based upon work supported by the National Science
Foundation (NSF) under Grants No. CNS-2312207, CNS-2107068, and CMMI-2246671. Any opinions, findings,
and conclusions or recommendations expressed in this material are
those of the authors and do not necessarily reflect the views of NSF.
\end{acks}

\bibliographystyle{ACM-Reference-Format}
\bibliography{reference}

\appendix

\section{Problem Formal Definition}
\label{sec:problemformal}

For clarity, we provide a formal definition of the scheduling problem as follows.

\noindent\textbf{Assumptions:}

\begin{itemize}
    \item One serverless function invocation uses at most one worker node.
    \item All worker nodes have the same hardware resources (we made this assumption for the explanation simplicity, our algorithm still works with heterogeneous hardware resources). 
    \item The serverless functions are in a form that can accept a single job or a batch of jobs.
\end{itemize}

\noindent\textbf{Given:} 
A set of jobs $B= \{ e_i \mid I \geq i>0\}$ with each job as an invocation of one of the serverless functions $\{f_j \mid J \geq j>0\}$; each job in B belongs to one and only one of the applications $A=\{a_m \mid M \geq m>0\}$ and one application consists of one or more jobs; each application $a_m$ has a tolerable latency upper limit $d_m$; a set of workers $\{ w_k \mid K \geq k>0\}$ with each worker being equipped with $R_C$ CPU resource units and $R_G$ GPU resource units; 

\noindent\textbf{Objective:} 
Produce a set of schedule configurations $C=\{c_l = (b_l, r_l, p_l, q_l) \mid L\geq l>0\}$, where each configuration $c_l$ represents that at time $q_l$, a set of jobs $b_l$ are assigned to the worker $p_l$ to run with $r_l$=($u_{c_l}$, $u_{g_l}$) resources ($u_c$ for CPU resource units, $u_g$ for GPU resource units), such that:

\begin{itemize}
    \item 
    $argmin_C \sum (\alpha u_{c_l} + \beta u_{g_l}), 1\geq \alpha \geq 0, \beta =1-\alpha$.
\end{itemize}

\noindent\textbf{Constraints:}
\begin{itemize}
    \item ${\sum h_m >\gamma}$, where $h_m$=1 if the end-to-end time of application $m$, denoted as $t_{a_m}$, is no greater than $d_m$, and 0 otherwise; $t_{a_m} = \sum t_{b_i}$ where $t_{b_i}$ is the end-to-end time of job $b_i$, that is, the time from its invocation to its completion.
    \item $\cup_l^L b_l = \cup_i^I e_i$ (every job is scheduled)
    \item $b_i \cap b_j = \emptyset$,\;\; $\forall l \geq i, j>0$  and $i \neq j$ (each job belongs to only one set)
    \item $\sum_{l=0}^{L}{u_{c_l}\mathbf 1(b_l, k,t)}\leq R_C,\;\; \forall t, \forall K\ge k> 0$, where, $\mathbf 1(b_l,k,t)$ is 1 if $b_l$ is active at time $t$ on worker $k$, and 0 otherwise. (within total CPU resource) 
    \item $\sum_{l=0}^{L}{u_{g_l}\mathbf 1(b_l, k,t)}\leq R_G,\;\; \forall t, \forall K\ge k> 0$, where, $\mathbf 1(b_l,k,t)$ is 1 if $b_l$ is active at time $t$ on worker $k$, and 0 otherwise. (within total GPU resource)
\end{itemize}

\section{ESG\_1Q algorithm}
\label{sec:pseudocode}

\begin{algorithm}
\caption{ESG\_1Q} 
\RestyleAlgo{ruled}
\KwIn {i $\gets$ the current stage}
\KwIn {\textbf{\em endStage} $\gets$ the final stage of the function sequence}
\KwIn  {\textbf{\em w} $\gets$ the longest waiting time; }
\KwIn  {\textbf{\em q} $\gets$ the time quota, which is got by the dominator-based-distribution method, of this sequence  }
\KwIn  {\textbf{\em The target latency ($G_{SLO}$) $\gets$ (SLO - w) $\times$ q }}
\KwIn {\textbf{\em ConfigLists[j]} $\gets$ the profiles of function $j$ sorted in increasing latency } 
\KwOut  {\textbf{\em configPQ}=\{\}, the final feasible configurations and sorting by the resource cost}
\KwData  {\textbf{\em minRSC} $\gets$ a sorted list to maintain K best paths, which is used for pruning on the resource usage. (\textbf{\em K} is the solution number we set)}
\KwData { \textbf{\em Paths}=\{\}, the feasible paths until now and sorting by the resources cost.}
\For{each \textbf{\em config} in \textbf{\em ConfigLists[i]}}{
    calculate \textbf{\em tLow, rscLow, recFastest}\;
    \If {\textbf{\em tLow} >= \textbf{\em $G_{SLO}$}}{
        break \Comment*[r]{Pruning on time}
    }
    \eIf{ \textbf{\em rscLow} >= \textbf{\em minRSC[K-1]}} {
        continue \Comment*[r]{Pruning on cost, minRSC[k-1] is the best\_full\_paths\_maxCost in the main paper} 
    }{
        Remove \textbf{\em minRSC[K-1]} and insert \textbf{\em rscFastest} into \textbf{\em minRSC}\;
        Add the \textbf{\em config} into \textbf{\em Paths}\;
    }
}
\While{\textbf{\em i}+1 <= \textbf{\em endStage}}{
    Reset \textbf{\em minRSC} = []\;
    \While{\textbf{\em Paths} is not empty}{
        Dequeue one \textbf{\em path} from \textbf{\em Paths}\;
        \For{each \textbf{\em config} in \textbf{\em ConfigLists[i+1]}}{
            \textbf{\em newPath} $\gets$ Extend the \textbf{\em path} by appending the \textbf{\em config}\;
            calculate \textbf{\em tLow, rscLow, recFastest}\;
            \If {\textbf{\em tLow} >= \textbf{\em $G_{SLO}$}} {
                break\;
            }
            \If{ \textbf{\em rscLow} >= \textbf{\em minRSC[K-1]}}{
                continue\;
            }{
                \If{\textbf{\em i}+1 == \textbf{\em endStage}}{
                    Insert the \textbf{\em newPath} into  the \textbf{\em configPQ}\;
                }{
                    Remove \textbf{\em minRSC[K-1]}, insert \textbf{\em rscFastest} into \textbf{\em minRSC}\;
                    Add \textbf{\em newPath} into \textbf{\em Paths}\;
               }
            }
        }
    }
    \textbf{\em i} = \textbf{\em i} + 1\;
}
\textbf{Return} \textbf{\em configPQ}\;
\label{alg:ESG1q}
\end{algorithm}










\end{document}